\documentclass{siamltex}
\usepackage[dvips]{graphics}
\usepackage{epsfig}

\newtheorem{remark}[theorem]{Remark}
\newtheorem{example}[theorem]{Example}

\DeclareRobustCommand{\qed}{%
  \ifmmode 
  \else \leavevmode\unskip\penalty9999 \hbox{}\nobreak\hfill
  \fi
  \quad\hbox{\qedsymbol}}
\newcommand{\openbox}{\leavevmode
  \hbox to.77778em{%
  \hfil\vrule
  \vbox to.675em{\hrule width.6em\vfil\hrule}%
  \vrule\hfil}}
\newcommand{\qedsymbol}{\openbox}
\renewenvironment{proof}[1][Proof.]{\par
  \normalfont\trivlist
  \item[\hskip\labelsep({\itshape#1})]\ignorespaces
}{\qed\endtrivlist}
\newcommand{\Fix}{\mbox{Fix}}
\newcommand{\e}{\varepsilon}
\title{Function Dynamics}
\date{\today}
\author{Yoichiro Takahashi\thanks{Research Institute for Mathematical
  Sciences, Kyoto University} \and
Naoto Kataoka\thanks{Meme Media Laboratory (VBL), Faculty of Engineering, Hokkaido University} \and
Kunihiko Kaneko\thanks{Department of Basic Science, Graduate School of Arts and Sciences, University of Tokyo} \and
Takao Namiki\thanks{Division of Mathematics, Hokkaido University}
}

\begin{document}
\maketitle
\pagestyle{myheadings}
\thispagestyle{plain}
\markright{Function Dynamics}

\begin{center}
Japanese Journal of Industrial and Applied Mathematics, 18-2, 2001
\end{center}


\begin{abstract}
We show mathematical structure of the function dynamics, i.e., the dynamics of interval maps 
$f_{n+1} = (1-\e)f_n + \e f_n\circ f_n$ and clarify the types of fixed points,
the self-referential structure and the hierarchical structure.
\end{abstract}

\section{Introduction}
Since about thirty years ago the late Professor Yamaguti had continued
 to recommend 
young researchers to find and study ``new mathematics in phenomena''.
Figure \ref{fig:1} below shows the phenomena we study in the present paper.

Let $I$ be an interval and for a given map $f:I\to I$
let us define a new map 
$\Phi_{\e}(f) : I\to I$ by 
\begin{equation}\label{eqn:1.1}
  \Phi_{\e}(f) = (1 - \e)f + \e f\circ f,
\end{equation}
where $0 < \e \le 1$.
Given $f_0:I\mapsto I$ we consider the function dynamics defined by
\begin{equation}
  f_{n+1} = \Phi_{\e}(f_n) \ (n = 0, 1, 2, \ldots).
\end{equation}

The original motivation to study (1.1) of [1] is
mentioned in Section 5, but the motivation of the present paper
consists in the study of (1.1) as an infinite dimensional
dynamical system.  For $\e =1$, $f_{n+1}= f_n \circ f_n$
is nothing but $2^n$-th iteration of the map $f_1$.  Therefore, one
might
expect very much complicated and chaotic behaviors in (1.1).
However, the simulations in [1,2] for $\e <1$ indicate
that (1.1) can exhibit rather simple behaviors with hierarchical and
self-referential structures, which we will
prove in a rigorous manner in the present paper.

The above dynamics can also be written as
\begin{equation}
f_{n+1} = g_n\circ f_n
\end{equation}
where $g_n$ is defined from $f_n$ by $g_n(x) = (1-\e)x + \e f_n(x)$.
The structure that $f_n$ gives $g_n$ and $f_{n}$ is evolved by $g_n$
 is a key to the emergence of what we call self-referential structure.

Figure \ref{fig:1} shows two typical examples of the phenomena observed in function
dynamics as $n\to\infty$ with $f_0(x) = rx(1-x)$.
In the simulation, we
take a finite mesh size to compute the function $f_n(x)$, although the ``phenomena''
we discuss is not an artifact of the finite mesh, but they remain as mesh points
are increased (or one can say that a piece-wise step function to approximate
$f_0(x)$  with a small mesh size).
As $n$ goes large in the simulation the flat parts of the graphs grow up rapidly and
they fill the whole interval in Figure \ref{fig:1}(a) (within 100 simulation steps
when mesh number = 4096). At each flat part $f_n(x)$ starts to be fixed in time
within some time steps. There appear finer 
flat parts with smaller intervals when the initial $r$ is larger or
$\e$ is smaller. Furthermore, 
there appear some complicated structure and some parts with irregular oscillation in time,
near the end points of flat parts in Figure \ref{fig:1}(b).
Those phenomena as well as other structures and dynamics were reported 
and heuristically analyzed in \cite{KKI} and \cite{KKII}.


\begin{figure}[htbp]
\begin{center}
  \includegraphics[width=5cm, height=5cm]{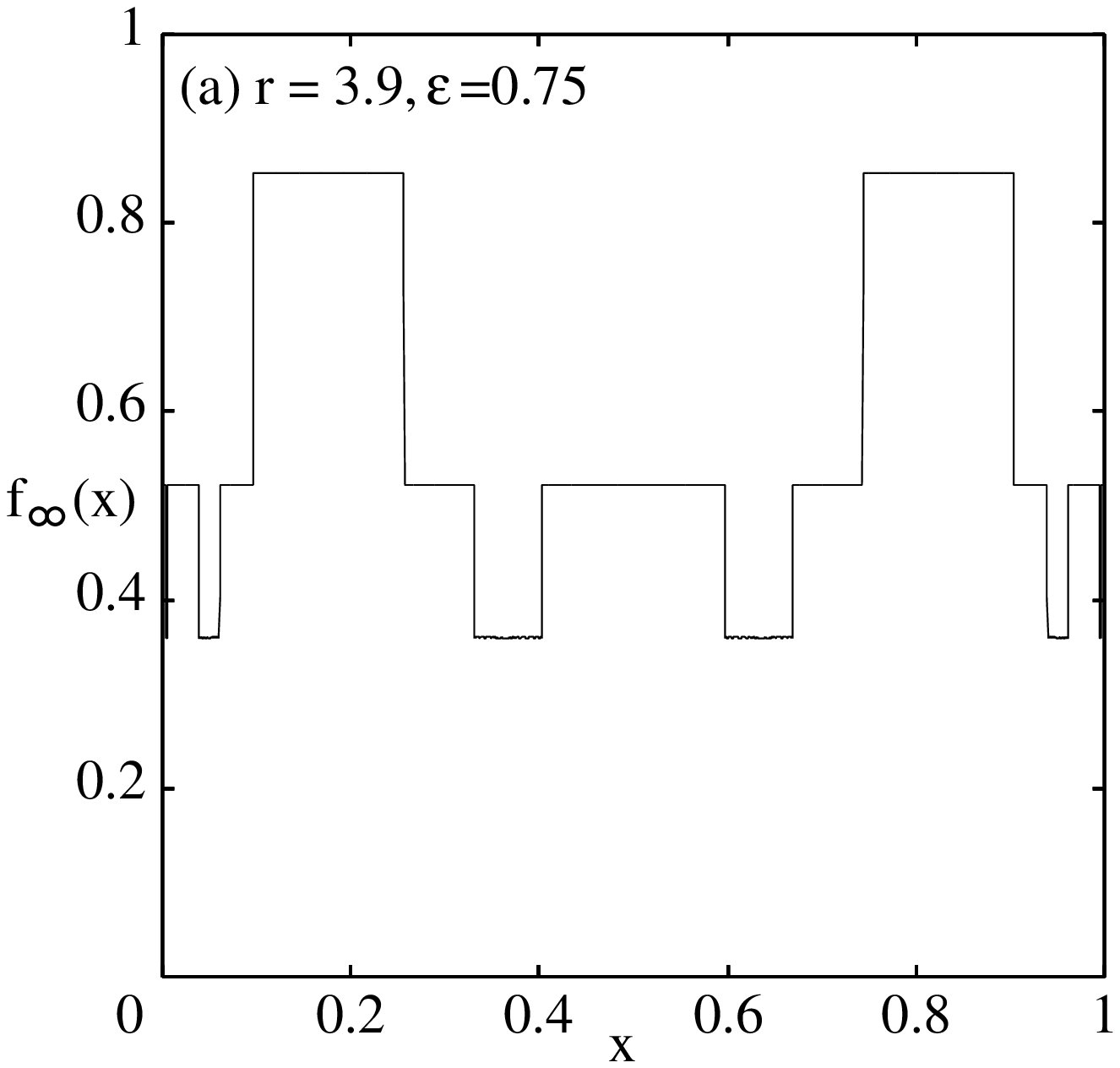}
  \includegraphics[width=5cm, height=5cm]{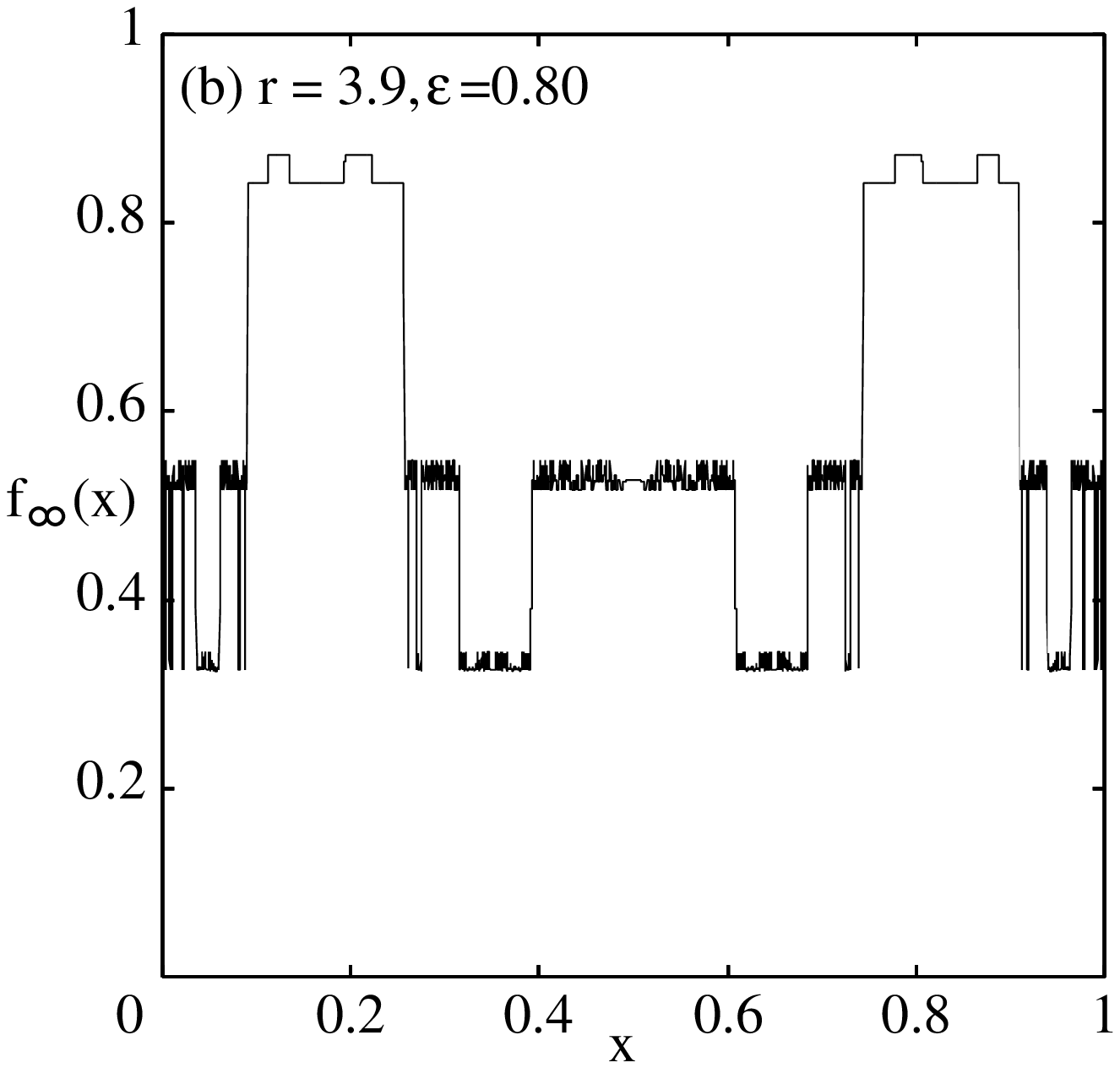}
  \caption{\label{fig:1} Two typical results $f_{\infty}(x)$ of numerical
simulation of function dynamics for $f_0(x) = rx(1-x)$. (a) $r = 
3.9$, $\e = 0.75$. (b) $r = 3.9$, $\e = 0.8$. $f_n(x)$ with sufficiently large
$n$ is plotted.  The simulation is carried out by using the
mesh number $4096$, i.e., by taking a piecewise step function approximating
$f_0(x)$ with the mesh size 1/4096.
}
\end{center}
\end{figure}

In the present paper,
we study the limit $f_\infty(x)$.  In Section 2,
the flat parts are explained.
In Section 3 we study how the ``self-reference'' is represented
within a hierarchical structure of the function $f_\infty(x)$.

In particular, there can exist trajectories such that 
$$f_{n+1}(x) =g_\infty(f_n(x))$$
where $g_\infty$ is the ``generated map'' in the terminology of \cite{KKI},
and some part of function is driven by other parts.
In Section 4, we give a further example which shows ``entangled hierarchy'',
where the hierarchy of rules change dynamically in time. 
In the last section  5, the original motivation of the model are discussed,
and the results are interpreted.

To close the introduction, we state some terminology.
We denote the set of fixed points of $f$ as $Fix(f)=\{x\in I|
f(x)=x\}$.  Take a fixed point $q\in Fix(f)$, and we call $q$ a
stable fixed point if there exists an open
neighborhood $U$ of $q$ such that $U\supset f(U)\supset \cdots\supset f^n(U)$ and
$\cap_{n\ge 0} f^n(U)=\{q\}$.  A semi-stable
fixed point $q$ is defined in a similar manner but $U$ has the form
$[q,q+\delta)$ or $(q-\delta,q]$.
 $Fix^{(s)}(f)$ denotes the set of
stable or semi-stable fixed points.  The basin of attraction $B(Q)$ of
$Q\subset Fix^{(s)}(f)$ is defined as $\{ x\in I| \lim_{n\to\infty}
f^n(x)\subset Q\}$. 

\section{The fixed point $f_{\infty}$}
Our starting point of the study is to focus on those points $x$ in the
interval $I$ where the limits 
\begin{equation}\label{eqn:1}
  f_{\infty}(x) = \lim_{n\to\infty} f_n(x)
\end{equation}
exist.

Optimists will take the formal limit of $f_{n+1} = (1-\e)f_n +
\e f_n\circ f_n$ to  
find the following relation independent of $\e > 0$:

\begin{equation}\label{eqn:2}
  f_{\infty}(x) = f_{\infty}(f_{\infty}(x)).
\end{equation}

In other words, it is expected that the limit $f_{\infty}(x)$ is, if any,
a fixed point of the map $f_{\infty}$
and that $f_{\infty}$ is a step function taking fixed point as its values.
Simulations support this.
Indeed, what we called the flat parts in Figure \ref{fig:1} form step functions.
Note, however, that domain of $f_{\infty}$ may not be the whole interval $I$.

The following is the mathematical statement for the above observation.

\begin{theorem}\label{theorem:1}
For a given continuous map $f_0:I\mapsto I$, there exist
a non-empty subset $\Omega$ of the interval $I$ and a map
$f_{\infty}:\Omega\mapsto\Omega$ which satisfy the following properties:

\begin{enumerate}
\item[(i)] For each $x\in\Omega$ the limit (\ref{eqn:1}) exists and
(\ref{eqn:2}) holds. 
\item[(ii)] $\Fix(f_{\infty})$ is non-empty.
\item[(iii)] $f_{\infty}(\Omega) \subset \Fix(f_{\infty})$. In other
  words, $f_{\infty}$ is a step function on $\Omega$ outside a (possibly empty) 
subset of $I$ where $f(x) = x$. More precisely,
  let $\Omega_q:=f_{\infty}^{-1}(q)=\{x\in\Omega| f_\infty(x)=q\}$, we have a
  partition:
  \[  \Omega = \bigcup_{q \in \Fix(f_{\infty})}\Omega_q.   \]
\end{enumerate}
\end{theorem}

\begin{remark}
   We should mention here that we do not exclude the case when the 
map $f_{\infty}$ is the identity map if it is restricted to a subinterval.
\end{remark}

 As the proof below shows, we can take
  \begin{equation}\label{eqn:3}
    \Omega = \bigcup_{n \ge 0}f_n^{-1}(\Fix(f_n)).
  \end{equation}
  In particular, $\Omega \neq \emptyset$ since 
  $\Fix(f_0) \neq \emptyset$ by the intermediate value theorem.
  Moreover, the set
  \begin{equation}\label{eqn:4}
    \Omega^I = \bigcup_{n\ge 0} \Fix(f_n)
  \end{equation}
  coincides with $\Fix(f_{\infty})$.
 In \cite{KKI}, the point in $\Omega^I$ is called the fixed point of
  type-I and the point in  $\Omega^{II} :=\Omega\setminus\Omega^I$ is
  called the fixed point of type-II.

	 The set $\Omega_q$ is an at most countable union of
    intervals if $q$ is a stable fixed point of some $f_n$, while
    $\Omega_q$ is a finite or at most countable set if $q$ is an
    unstable fixed point. 
The following lemma guarantees that flat parts of the graph of $f_n$ grow up.

\begin{lemma}\label{lemma:1}
\begin{enumerate}
  \item[(i)] $\Fix(f_n)\subset\Fix(f_{n+1})$ for each $n$.
  \item[(ii)] $f_n^{-1}(\Fix(f_n))\subset
    f_{n+1}^{-1}(\Fix(f_{n+1}))$ for each $n$.  
\end{enumerate}
\end{lemma}

\begin{proof}
If $f_n(x) = x$, then $f_{n+1}(x) = (1-\e)f_n(x) + \e
f_n(f_n(x)) = (1-\e)x + \e x = x$. 

Hence, $x\in \Fix(f_{n+1})$.

Next, if $f_n(x) = q$ and $f_n(q) = q$
then $f_{n+1}(q) = q$ by (i) and 
$f_{n+1}(x) = (1 - \e)f_n(x) + \e f_n(f_n(x)) = (1-\e)q + \e f_n(q) = q$.

Hence, $x \in f_{n+1}^{-1}(\Fix(f_{n+1}))$.
\end{proof}

\begin{proof}[Proof of Theorem \ref{theorem:1}]
Define the sets $\Omega$ and $\Omega^I$ by (\ref{eqn:3}) and (\ref{eqn:4}).
If $x\in \Omega^I$, then $x \in \Fix(f_n)$ for some $n$,
and therefore, by lemma \ref{lemma:1}(i), $f_n(x) = x$ for all $m \ge n$.
Hence, the limit $f_{\infty}(x)$ exists and equals $x$.
If $x \in \Omega$, then $x \in f_n^{-1}(\Fix(f_n(x))$ for some $n$.
By Lemma \ref{lemma:1}(ii) (and its proof)
\[
  f_m(x) = q \mbox{ for all } m \ge n \mbox{ with } q = f_n(x) \in \Fix(f_n).
\]
Hence, the limit $f_{\infty}(x)$ exists and equals $q$.
In particular, $f_{\infty}(f_{\infty}(x)) = f_{\infty}(q) = q = f_{\infty}(x)$.
Consequently we obtain (i) and (ii).
Now (iii) follows if we set
\[
  \Omega_q = f_{\infty}^{-1}\{q\} = \{x \in \Omega | f_{\infty}(x) = q\}.
\]
\end{proof}

The case where $f_{\infty}$ is a continuous function is very restrictive:

\begin{proposition}
If $f:I\to I$ is continuous and satisfies $f\circ f = f$ on $I$,
then $\Fix (f)$ is an interval and $I = f^{-1}\Fix (f)$.
\end{proposition}

\begin{proof}
The condition $f\circ f = f$ on $I$ implies $I = f^{-1}\Fix (f)$.
Suppose the contrary. Then we would find two fixed points $q_0$ and $q_1$ and 
a non fixed point $x$ in between $q_0$ and $q_1$.
Let $q_0 < x < q_1$. Since $f$ is continuous,
$f[q_0, q_1] \supset [q_0, q_1]$. Hence the intermediate value theorem implies that 
there exist a point $y$ in $(q_0, q_1)$
such that $f(y) = x$.
Then we would have $f(f(y)) = f(x) \neq x$,
which contradicts $f\circ f = f$.
\end{proof}


\begin{figure}[htbp]
  \begin{center}
    \includegraphics[width=6cm, height=6.5cm]{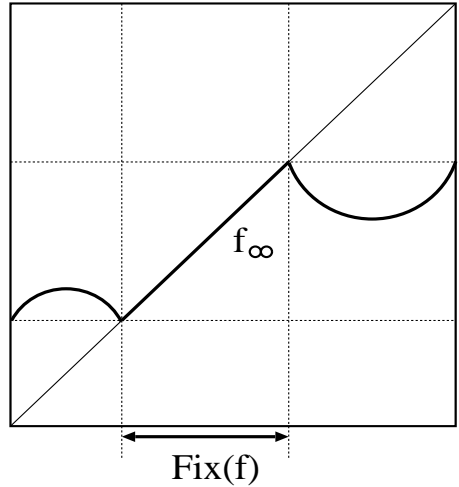}
  \end{center}
  \caption{\label{confix:fig} The case $f_{\infty}$ is continuous.}
\end{figure}

\begin{figure}[htbp]
  \begin{center}
    \includegraphics[width=7cm, height=7cm]{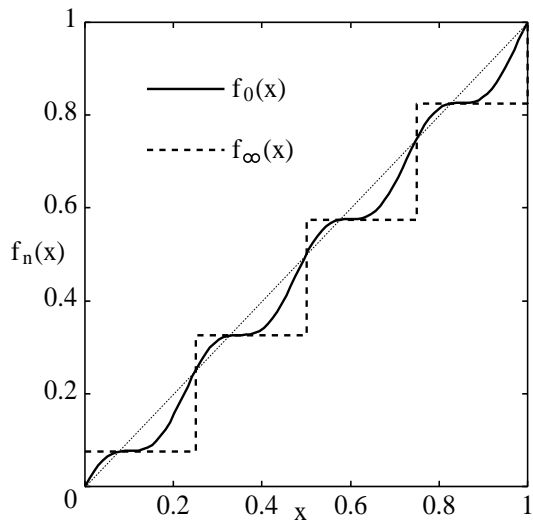}
  \end{center}
  \caption{\label{fig:2} The evolution of the function when $f_0$ is monotone 
nondecreasing and bounded. $f_0(x)$ (solid line) 
and $f_{\infty}(x)$ (dotted line) are plotted.}
\end{figure}

An example of such a $f_{\infty}$ is shown in Figure \ref{confix:fig}.
The continuous $f_{\infty}$ is very restrictive,
because even in a very simple case the function $f_{\infty}$ is a non-continuous step function.
The relevant $f_{\infty}$ is actually a step function.
Simple examples are as follows:

\begin{example}
\label{ex:mono}
If $f_0(x)$ is monotone nondecreasing and continuous,
then $\Omega = I$ and $\Fix^{(s)}(f_0)\subset \Omega^I\subset \Fix(f_{\infty}) = \Fix(f_0)$ 
(See Figure \ref{fig:2}). 
\end{example}

\begin{proof}
Let $J$ be a connected component of the set $\{x \in I | f_0(x) > x\}$.
Then, $f_0(J)\subset J$ by the assumption on $f_0$.

Thus, if $x\in J$ then the sequence $f_0(x), f_0(f_0(x)), \ldots$ is
monotone nondecreasing and bounded 
and so it has a limit $f_{\infty}(x)$, which is necessarily a fixed
point of $f_0$. 
In particular, the subsequence $\{f_n(x)\}_{n=0,1,2,\ldots}$
converges to a fixed point of $f_0$. 

Similarly $f_n(x)$ converges to a fixed point of $f_0$
for any $x$ is a connected components of $\{x\in I| f_0(x) < x\}$
\end{proof}


\begin{example}
\label{ex:stable}
If the continuous map $f_0: I\mapsto I$ has a stable fixed point
whose basin of attraction
is $I$ or coincides with $I$ except for an unstable fixed point, then
$\Omega = I$ and  
$\Omega^I = \Fix(f_\infty) = \Fix(f_0)$.
\end{example}

Now we consider the stable fixed points.

\begin{lemma}\label{lemma2.6}
If $q$ is a (semi-)stable fixed point of $f_0$, then $q$ is a
(semi-)stable fixed point of every $f_n$, $n\ge 0$.
\end{lemma}

\begin{proof}
By definition, one can take a semi-open interval $U$ which is either
of the form $U=(q-\delta,q]$ or $[q,q+\delta)$ with $\delta>0$,
$U\supset f_0(U)\supset\cdots\supset f_0^k(U)\to\{q\}$ as $k\to\infty$
and $f_0$ is one-to-one on $U$.  Then,
\begin{eqnarray*}
  f_1(U)& \subset & (1-\e)f_0(U)+\e f_0(f_0(U))\\
	& \subset & (1-\e)f_0(U)+\e f_0(U)\subset f_0(U)\subset U
\end{eqnarray*}
where $(1-\e)A+\e B :=\{(1-\e)a+\e b| a\in A,\ b\in B\}$. 

Similarly, for each $k \geq 1$, $f_1(f_0^k(U)) \subset f_0^{k+1}(U) \subset f_0^k(U)$.

Hence, we get 
\[
  U \supset f_1(U)\supset\cdots\supset f_1^k(U)
\]
and on the other hand
\[
  f_1^k(U)\subset f_0^k(U).
\]
Consequently, 
\[
  U\supset f_1(U)\supset\cdots\supset f_1^k(U)\to\{q\}.
\]
The assertion for $n\ge 2$ follows by induction.
\end{proof}

Now we discuss the degree of stability of fixed points.  Let $q$ be a fixed point
of $f_n$, $U$ be an interval contain $q$.  Set
\[ 
  l_n(x;q)=l_n(x)=\frac{f_n(x)-q}{x-q}\ \ \mbox{for $x\in U$ and
  $x\ne q$}.
\]
Then, 
\begin{equation}
l_{n+1}(x)= (1-\e)l_n(x)+\e l_n(f_n(x))l_n(x).
\end{equation}
Define $\rho_n=\rho_n(U)$ by $\rho_n=\sup_{x\in U,x\ne q}|l_n(x)|$, we get
\begin{equation}\label{eqn:stability}
  0 \le \rho_{n+1}\le (1-\e)\rho_n+\e\rho_n^2.
\end{equation}

This inequality shows that the dynamical system $h(x)=(1-\e)x+\e x^2$
controls the stability of fixed points of $f_n$.  From this
observation we can show the following. 

\begin{lemma}\label{lemma2.7}
Let $q$ be a (semi-)stable fixed point of $f_0$ and $U_q$ be a semi-open interval $U_q$ which is either
of the form $U_q=(q-\delta,q]$ or $[q,q+\delta)$ with $\delta>0$,
$U_q\supset f_0(U_q)\supset\cdots\supset f_0^k(U_q)\to\{q\}$ as $k\to\infty$
and $f_0$ is one-to-one on $U$.
If $\rho_0(U_q)<1$, then $\lim_{n\to\infty}f_n(x)=q$.
\end{lemma}

\begin{proof}
If $f_n(x)=q$ for some $n$, $\lim_{n\to\infty}f_n(x)=q$. 

Suppose $f_n(x)\ne q$ for all $n$, then
\begin{eqnarray*}
|f_{n+1}(x)-q|&=&|(1-\e)(f_n(x)-q)+\e (f_n(f_n(x))-q)|\\
  &=& \left|(1-\e)(f_n(x)-q)+\e
  (f_n(x)-q)\frac{f_n(f_n(x))-q}{f_n(x)-q}\right|\\
  &=& |f_n(x)-q|\cdot |(1-\e) + \e l_n(f_n(x))|
\end{eqnarray*}

By inequality \ref{eqn:stability} and $\rho_0<1$,
$$|l_n(f_n(x))|<\rho_n < h(\rho_{n-1}) < h^n(\rho_0)\to 0 \mbox{ as
$n\to\infty$}.$$

As a result, there exist $\delta$ such that $0\le |(1-\e) + \e
l_n(f_n(x))| < \delta < 1$ for large $n$ and we have $|f_{n+1}(x)-q| <
\delta |f_n(x)-q|$.  This implies $\lim_{n\to\infty}f_n(x)=q$.
\end{proof}

\begin{lemma}\label{lemma2.8}
  For $x\in f_0^{-1}(U_q)$, $\lim_{n\to\infty}f_n(x)=q$.
\end{lemma}

\begin{proof}
In the proof of lemma \ref{lemma2.7}, a key inequality is
$|l_n(f_n(x))| < \rho_n$ for $f_n(x)\in U_q$.
Now $f_0(x)\in U_q$ from assumption.  So $f_n(x)\in U_q$ by induction.
The proof is similar.
\end{proof}

  From Lemmas \ref{lemma2.6}-\ref{lemma2.8}, we can extend Theorem
\ref{theorem:1} as follows: 

\begin{proposition}\label{lemma:3}
For each $n$ and $q\in Fix^{(s)}(f_n)$ take any interval $V_q$
where $f_n$ is monotone and such that $\rho_n(V_q)<1$. Set
\[\Omega=\bigcup_{n\ge 0}f_n^{-1}\left(\bigcup_{q\in\Fix^{(s)}(f_n),
  }V_q\right).\]
Then the set $\Omega$ satisfies the conditions (i), (ii) and (iii)
in Theorem \ref{theorem:1}.
\end{proposition}

\begin{figure}[htbp]
  \begin{center}
  \includegraphics[width=7cm, height=5cm]{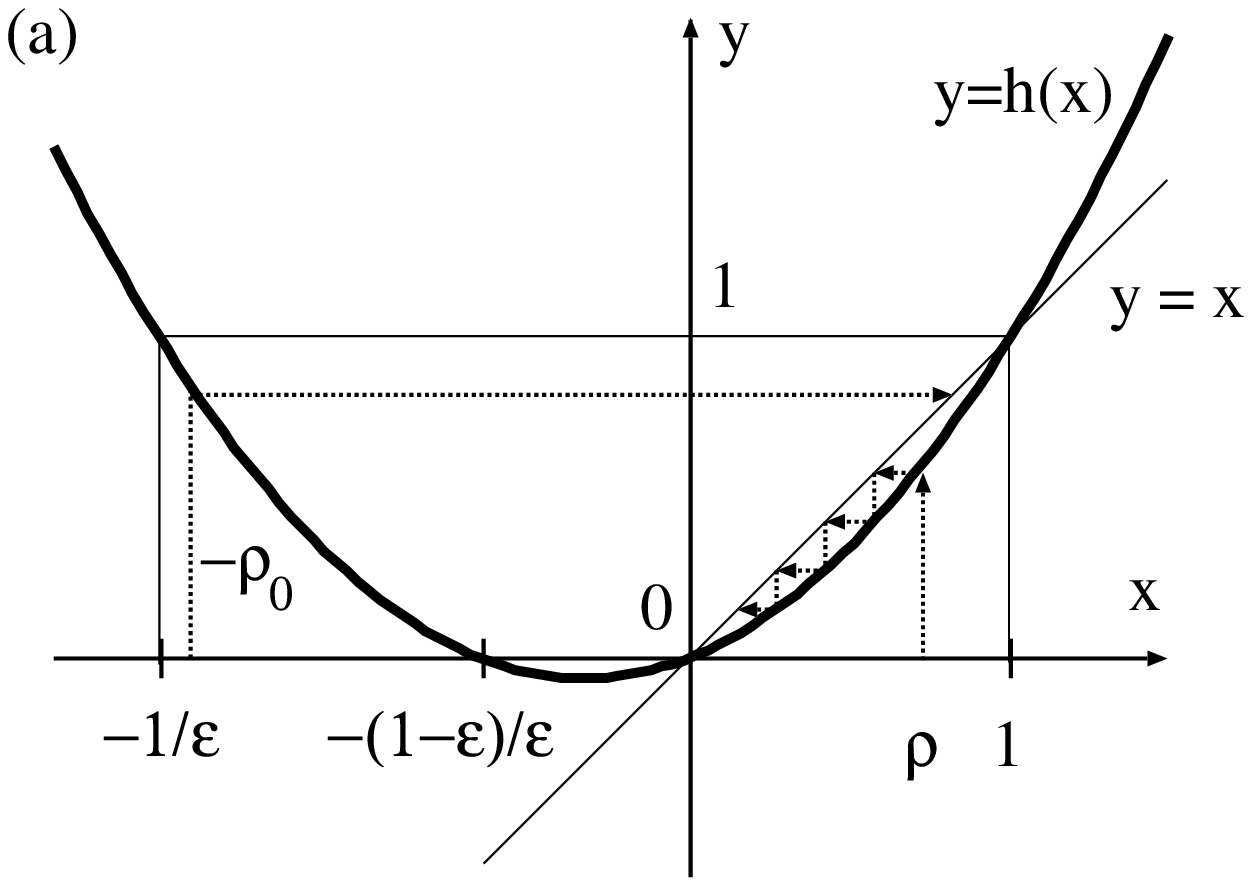}
  \includegraphics[width=5cm, height=5cm]{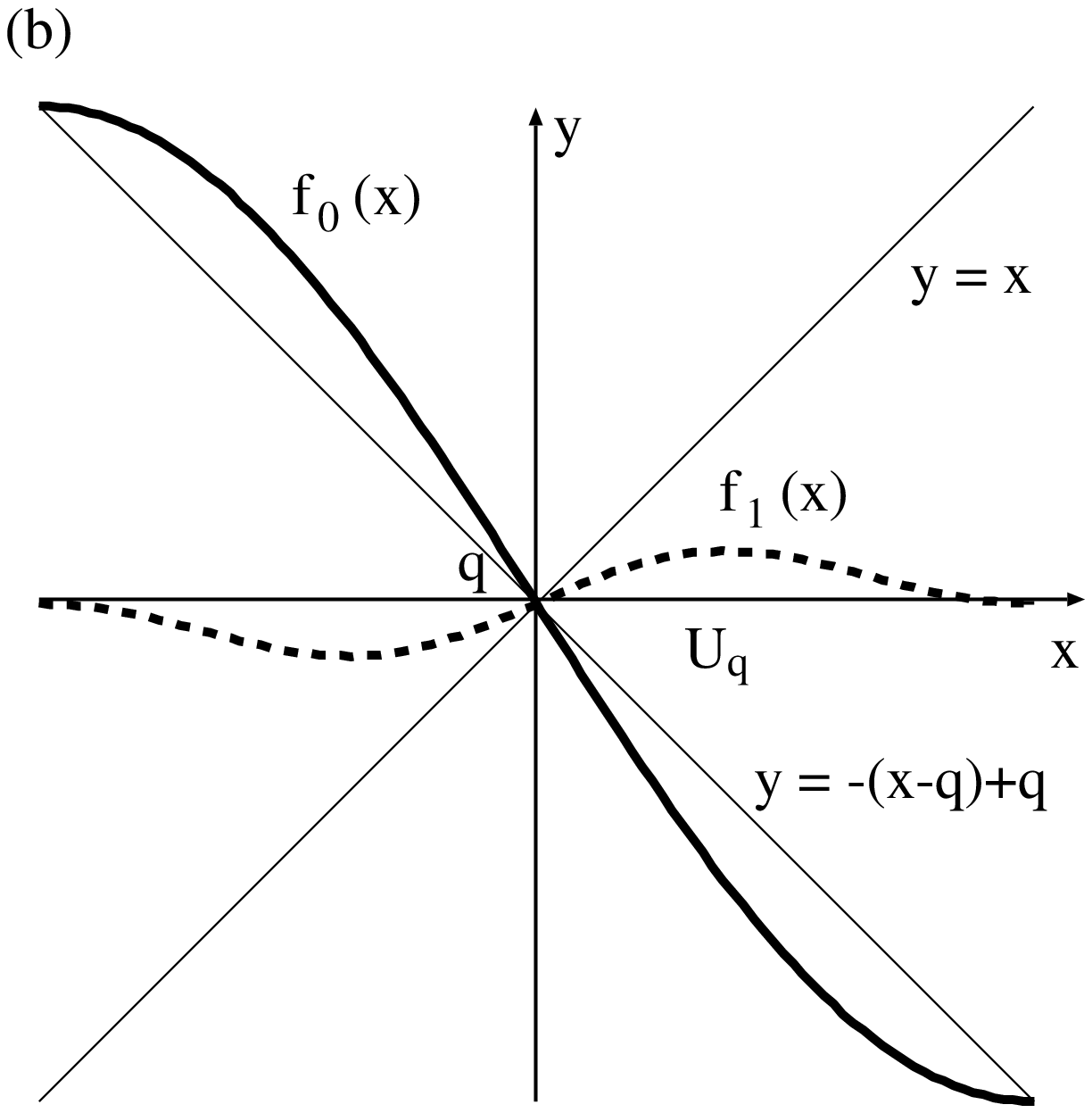}
  \end{center}
  \caption{(a) Dynamics of $\rho_n$ and its example. (b) $f_0(x)=-\sin(\pi/2 x)$ and $\e=0.5$.}
  \label{fig:stability}
\end{figure}

\begin{remark}
  There exist some unstable fixed points which become stable after iteration. 
  For example, let $q$ be a unstable fixed point of $f_0$ and assume
  that $f_0$ be monotone decreasing on $U_q$, $U_q\supset f_0(U_q)$
  and $-1/\e <-\rho_0<-(1-\e)/\e$. 
  Then $q$ becomes a stable fixed point of $f_1$ because $\rho_1<1$ as
  shown in Figure \ref{fig:stability}.
\end{remark}


\section{Trajectories driven by $g_{\infty}$ and Hierarchical Map}
\label{sec:3}
Let us study the asymptotic behavior of trajectories other than
$(\Omega, f_{\infty})$. 
For this purpose, as is mentioned in Introduction, we write
\begin{equation}\label{eqn:g:1}
  f_{n+1} = g_n\circ f_n
\end{equation}
by setting
\begin{equation}\label{eqn:g:2}
  g_n(x) = (1 - \e)x + \e f_n(x).
\end{equation}
Since our target is the asymptotic behavior as $n\to\infty$,
we may assume that $f_n|_\Omega$ is close to $f_{\infty}$ from the beginning.
For simplicity, we assume
\begin{equation}\label{eqn:g:3}
  f_0|_\Omega = f_{\infty}.
\end{equation}

\begin{lemma}\label{lemma:2}
If (\ref{eqn:g:3}) holds, then
\begin{equation}\label{eqn:g:4}
  f_n|_\Omega = f_{\infty}
\end{equation}
for all $n = 0, 1, 2, \ldots$
\end{lemma}

\begin{proof}
Assume $x\in\Omega$ and $f_n(x) = q \in \Fix(f_{\infty})$.  Then,
\[
  f_{n+1}(x) = (1-\e)q + \e f_n(q) = q = f_n(x) = f_{\infty}(x).
\]
Hence, $f_n|_\Omega = f_{\infty}$ implies $f_{n+1}|_\Omega = f_{\infty}$
and (\ref{eqn:g:4}) follows by induction on $n$. 
\end{proof}

By Lemma \ref{lemma:2}, the ``generated map'' $g_n$ also coincides
with $g_{\infty}$, if it is restricted to $\Omega$: 
\begin{equation}\label{eqn:g:6}
g_{\infty}(x) = (1 -\e)x + \e q \ \mbox{ if $x\in \Omega_q$}.
\end{equation}

\begin{example}[Nagumo-Sato map]
\label{ex:NS}
Let the initial map $f_0:I\to I$, $I\supset [\frac{a - 1}{\e},
\frac{a}{\e}]$ be as follows:
\[
  f_0(x) = 
  \left\{
  \begin{array}{cl}
  \frac{a}{\e},& x \in \{\frac{a}{\e}\}\cup (\frac{a - 1}{\e}, 0),\\
  \frac{a-1}{\e},& x \in [0,\frac{a}{\e})\cup\{\frac{a-1}{\e}\},\\
  b(x),& x \in I \setminus [\frac{a - 1}{\e}, \frac{a}{\e}]. 
  \end{array}
  \right.
\]
Here $0 < a < 1$ and $b(x)$ is a function satisfying a condition 
$b(x)\in(\frac{a-1}{\e},\frac{a}{\e})$.  Then
\[
  \Omega = \left[\frac{a-1}{\e}, \frac{a}{\e}\right]
\]
and
\[
  f_{\infty}(x) = 
  \left\{
  \begin{array}{cl}
  \frac{a}{\e}, &x\in\{\frac{a}{\e}\}\cup(\frac{a - 1}{\e},0),\\
  \frac{a-1}{\e}, & x \in [0, \frac{a}{\e}) \cup \{\frac{a-1}{\e}\}.
  \end{array}
  \right.
\]
\end{example}

\begin{figure}[htbp]
 \begin{center}
    \includegraphics[width=6cm,height=6cm]{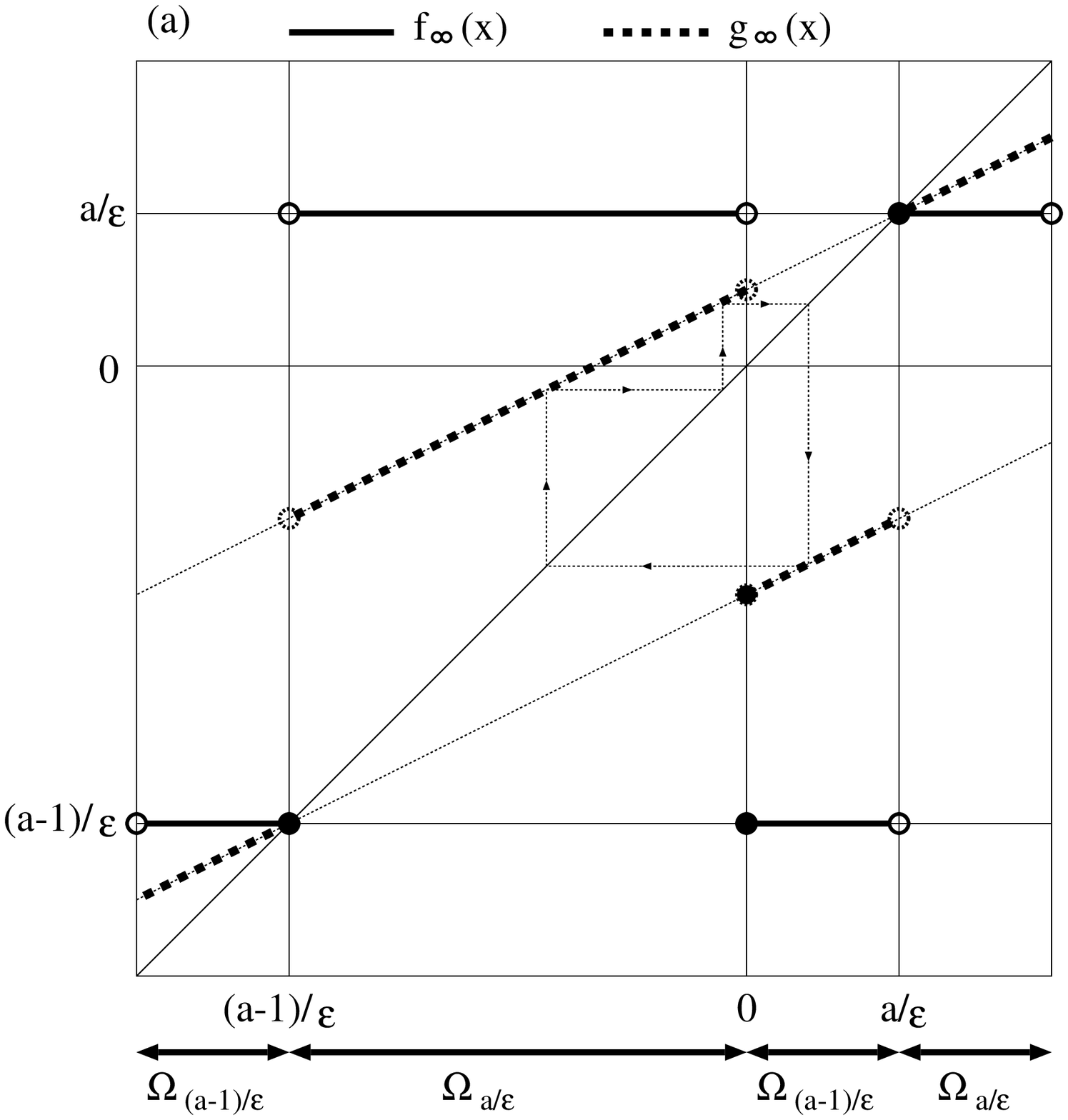}
    \includegraphics[width=6cm,height=6cm]{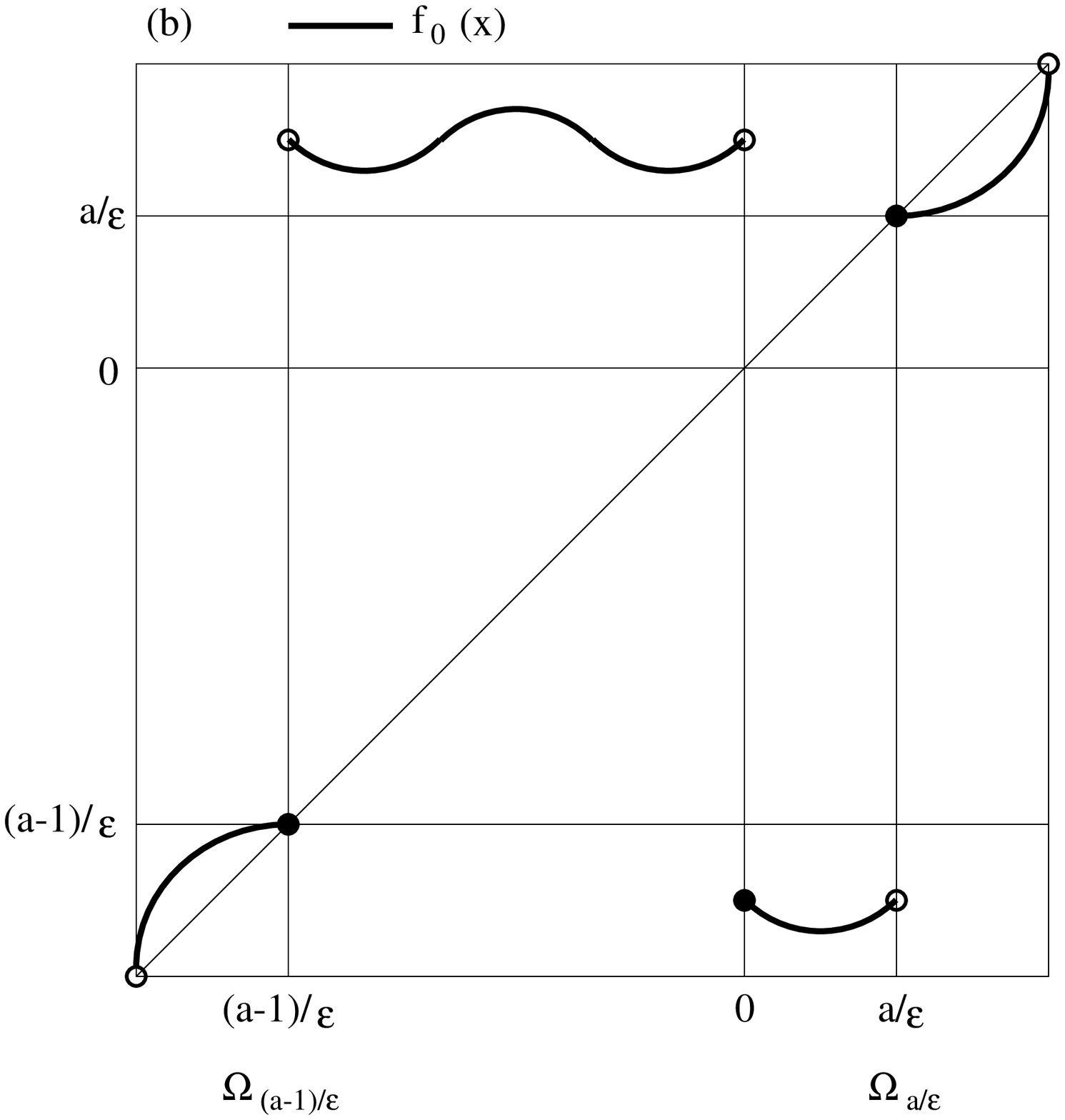}
    \caption{\label{gooNS} (a) The graph of $f_{\infty}(x)$ (solid line) and the
    generated map $g_{\infty}(x)$ (dotted line). Here $\e = 1/2$ and $a =
    1/4$. The map $g_{\infty}$ has a period-3 attractor. (b) An
    example of a graph of initial function $f_0(x)$ which converges to
    $f_{\infty}$ displayed in (a).} 
 \end{center}
\end{figure}

For instance suppose $f_0(x)$ is given by 
the Figure \ref{gooNS} (b). Then $f_n$ converges to the $f_{\infty}$ shown in Figure \ref{gooNS}(a).
In this case, the generated map $g_{\infty}(x)$ is a piecewise linear
map defined by:
\[
  g_{\infty} = \left\{
    \begin{array}{cl}
       (1-\e)x + a , & x \in \{\frac{a}{\e}\}\cup (\frac{a - 1}{\e}, 0),\\
       (1 - \e)x + a-1,& x\in [0,\frac{a}{\e}) \cup \{\frac{a-1}{\e}\}
    \end{array}\right.
\]
(See Figure \ref{gooNS}(a)).


In this case, $g_{\infty}(\Omega) \subset \Omega$ and the map $g_{\infty}:\Omega \mapsto \Omega$
is studied in \cite{NS} and is called Nagumo-Sato map.
It is shown that a periodic orbit of any period can appear if one controls $\e$ and $a$ suitably.
Moreover, Cantor attractors (almost periodic orbits) can appear.

Combining (\ref{eqn:g:1}), (\ref{eqn:g:2}) and (\ref{eqn:g:6}) we see
$  f_{n+1}(x) = g_{\infty}(f_n(x)) \ \mbox{if} \ f_n(x) \in \Omega$.
Thus we obtain the trajectories $f_n(x)$ in $\Omega$, which is driven
by the Nagumo-Sato map $g_{\infty}$.

The above Example \ref{ex:NS} is generalized to the following.
\begin{theorem}
Assume $f_0|_\Omega = f_{\infty}$.
Define $g_{\infty}:\Omega\mapsto I$ by (\ref{eqn:g:6}). Set
\begin{equation}\label{eqn:g:12}
  \Omega (g_{\infty}) = \bigcap_{n \ge 0}g_{\infty}^{-n}(\Omega) = \{x
  \in \Omega | \ g_{\infty}(x), g_{\infty}(g_{\infty}(x)), \ldots \in
  \Omega\} 
\end{equation}
and assume $\Omega (g_{\infty})$ is non-empty.
Then, the trajectories starting from $f_0^{-1}(\Omega (g_{\infty}))$ are ``driven'' by $g_{\infty}$.
Precisely, if $x \in I$ and $f_0(x) \in \Omega (g_{\infty})$ then
\begin{equation}\label{eqn:g:13}
  f_n(x) \in \Omega (g_{\infty}) \ \mbox{for all} \ n = 0, 1, 2, \ldots
\end{equation}
and
\begin{equation}\label{eqn:g:14}
  f_n(x) = \underbrace{g_{\infty}\circ \cdots \circ g_{\infty}}_{\mbox{$n$ times}}(f_0(x)). 
\end{equation}

In other words, the trajectory $\{f_n(x)\}_{n = 0, 1, 2, \ldots}$ on $\Omega(g_{\infty})$ is reduced to the
$g_{\infty}$-orbit of $f_0(x)$.
\end{theorem}

\begin{proof}
If we show (\ref{eqn:g:13}), then (\ref{eqn:g:14}) will be obvious
from (\ref{eqn:g:6}). 
Assume $f_n(x) \in \Omega (g_{\infty})$, then $f_{n+1}(x) =
g_{\infty}(f_n(x))$.

By the definition (\ref{eqn:g:12}),
$g_{\infty}(f_n(x)) \in g_{\infty}(\Omega (g_{\infty})) \subset \Omega$. 

Hence $f_{n+1}(x) \in \Omega (g_{\infty})$.
\end{proof}

Now we proceed to the next stage and seek for the set $\Psi$ such that $f_n(\Psi) \subset \Omega$
and $g_n(\Psi)\subset \Psi$ for all $n$.

%
%

\begin{theorem}\label{theorem:3}
Assume that $f_0|_\Omega =f_\infty$ and that there exist non empty
$f_\infty$-invariant subsets $\Omega_i$, $i=1, \ldots, N,$ of $\Omega(g_{\infty})$ and closed 
subsets $\Psi_i$, $i=1, \ldots, N,$ of 
the complement of $\Fix(f_\infty)$ which satisfies the following property:
For each $i$ there exists a $j$ such that
\begin{equation}\label{eqn:5}
  (1-\e)\Psi_i +\e\Omega_i\subset\Psi_j.
\end{equation}
Let $X_i =f_0^{-1}(\Psi_i)$. If $f_0(\Psi_i)\subset\Omega_i$ and
$g_0(\Omega_i)\subset\Omega_i$, then
$g_0(\Psi_i)\subset\Psi_j$. Moreover, for any $n\ge 1$, 
\begin{equation}\label{eqn:6}
  f_n(X_i)\subset \Psi_i, f_n(\Psi_i)\subset \Omega_i
  \mbox{ and } g_n(\Psi_i)\subset\Psi_j.
\end{equation}

Hence, once a trajectory $f_n(x)$ falls into some $\Psi_i$, say, $f_{n_0}(x)\in \Psi_i$,
the trajectory $f_n(x)$, $n = n_0+1, n_0+2,\ldots$ is confined in $\bigcup \Psi_j$ and 
driven by $g_n$'s:
$f_{n+1} = g_n(f_n(x))$ for $ n = n_0+1, n_0+2, \ldots$.
\end{theorem}


\begin{proof}
By (\ref{eqn:5}) and the definition of $g_0$,
\[
  g_0(\Psi_i)\subset (1-\e)\Psi_i + \e f_0(\Psi_i)\subset 
  (1-\e) \Psi_i+\e\Omega_i\subset\Psi_j.
\]
Now, $f_1(X_i)=g_0(f_0(X_i))\subset g_0(\Psi_i) \subset \Psi_j$ and
$f_1(\Psi_i)=g_0(f_0(\Psi_i))\subset g_0(\Omega_i)$.
By Lemma \ref{lemma:2}, $g_0|_\Omega = g_\infty|_\Omega$.
Thus $g_0(\Omega_i)\subset\Omega_i$. 
Hence (\ref{eqn:6}) follows by induction.
\end{proof}

\begin{remark}
\label{remarkhi}
If there exist closed subsets $\{X_i\}_{i = 1}^M$ such that
\begin{equation}
  (1-\e)X_i +\e\Psi_i\subset X_j \ \ \mbox{for some $j$},
\end{equation}
the dynamics of $f_n|_{\Phi_i}$ ($f_n(\Phi_i)\subset X$) is determined by $g_n|_X$.
This process can continue ad infinitum and
it is not difficult to extend the Theorem \ref{theorem:3}.
\end{remark}

In \cite{KKII} the generated map $g_n|_{\Psi}$ is called the meta-map,
taking into consideration the point that
the dynamics of $g_n|_{\Phi}$ is determined
by $g_\infty$. Similarly, in \cite{KKII},
the generated map $g_n|_{X_i}$ is called meta-meta-map,
while the generated map $g_n$ is called hierarchical map as a whole.


Now we present two typical examples.  

The first example (Example 3.6)  shows typical
trajectories driven 
by $g_\infty$.  On $\Psi_i$ there are two branches of $f_n$,
$f_{even}$ and $f_{odd}$ while on $\Omega_i$ $f_\infty$ exists.
The dynamics of $f_n|_{X_i}$ is determined by $g_n|_{\Psi_j}$.

The second example illustrates the case of $f_n$ with further two branches on $X_i$.
$f_n|_{\Psi}$ is driven by $g_{\infty}$, while $f_n|_{X}$ is driven by $g_n|_{\Psi}$, and 
$f_n|_{\Phi}$ is driven by $g_n|_{X}$ hierarchically.
Each partial function is period 2 or time-invariant.
The configuration of initial function $f_0$ is given by 
nesting the initial function of the first Example \ref{ex:4}.\\

\begin{example}[meta-map]
\label{ex:4}
In this example, a new initial function which generates a meta-map is shown.
This initial function is given by a `surgery' of the $f_{\infty}$ which generates a map having a 2-period attractor.
The partial function $f_n|_{\Psi}$ ($f_n(\Psi)\subset \Omega$) is set to generate a 
time-dependent $g_n|_{\Psi}$ which has another period-2 attractor (meta-map).

Let $f_0(x)$ be as follows;
\[
  f_0(x) =
  \left\{
  \begin{array}{rll}
  -(a+b), & x \in \Omega_{-(a+b)}, &\Omega_{-(a+b)}:= \{-(a+b)\}\cup (-a, -(a-b)),\\
  -(a-b), & x \in \Omega_{-(a-b)}, &\Omega_{-(a-b)}:= \{-(a-b)\}\cup (-(a+b), -a],\\
  a-b,    & x \in \Omega_{a-b},    &\Omega_{a-b}:= \{a-b\}\cup (a, a+b),\\
  a+b,    & x \in \Omega_{a+b},    &\Omega_{a+b}:= \{a+b\}\cup (a-b, a],\\
       & &\\
   -a+Eb, & x\in \Psi_0, &\Psi_0:= (0, a-b),\\
    a+Eb, & x\in \Psi_1, &\Psi_1:= (-(a-b), 0],\\
          &  &          \\
  -E(a+Eb), & x \in X_0, &\\
   E(a-Eb), & x \in X_1. &\\
  \end{array}
  \right.
\]
Here $E := \frac{\e}{2-\e}$ and $b = \frac{1-\e}{1+\e} \cdot a$ ($a > 0$).
The graph of this initial function is shown in Figure \ref{meta1}.
The initial function $f_0$ on $\Omega_i$ is similar to the $f_0$ with two fixed points
in Example \ref{gooNS}, two copies of which are now embedded in subintervals
$[-(a+b),-(a-b)]$ and $[a-b, a+b]$ for the initial function $f_0$.
Here, the function which generates a map $g_{\infty}$ having period-2 attractor is embedded to the subintervals.
Now, $\Fix (f_0) = \{\pm (a+b), \pm (a-b)\}$.

Now,
(i) The generated map $g_{\infty}$ has a period-2 attractor.
(ii) $f_0|_{\Psi}$ ($f_n(\Psi) \subset \Omega$ for all $n$) is on the attractor of $g_{\infty}$. 
(iii) The $f_0|_{\Psi}$ is arranged so as to generate a time-dependent map $g_n|_{\Psi}$ (meta-map),
which has another period-2 attractor.
(iv) $f_0|_X$ ($f_n(X) \subset \Psi$ for all $n$) is on the attractor of $g|_{\Psi}$.
Each partial function is already on one of the attractors and $f_n$ is a period-2 function 
as a whole.

The procedure of time evolution is demonstrated straightforwardly as follows
(For the computation of each step, it is convenient to use 
the relation $(1-\e)Ea + \e(-a) = -Ea$).\\

At $n = 0$, the following conditions are satisfied.
\[
\left\{
  \begin{array}{ll}

  f_0(\Psi_0) \subset \Omega_{-(a+b)}, & \\
  f_0(\Psi_1) \subset \Omega_{a-b}, &\\
                               &    \\
  f_0(X_0)    \subset \Psi_1,   &\\
  f_0(X_1)    \subset \Psi_0.   &\\

  \end{array}
  \right.
\]

At the next step, this $f_0(x)$ evolves to the following $f_1$;

\[
\left\{
  \begin{array}{llrr}

  f_1|_{\Psi_0} &= 
(1-\e)(-a+Eb) + \e(-(a+b)) &=& -a-Eb,\\
  f_1|_{\Psi_1} &= 
(1-\e)(a+Eb) + \e(a-b) &=& a-Eb,\\
                               &    &\\
  f_1|_{X_0}    &= 
(1-\e)(-E(a+Eb)) + \e (a+Eb) &=& E(a+Eb),\\
  f_1|_{X_1}    &= 
(1-\e)(E(a-Eb)) + \e (-a+Eb) &=& -E(a-Eb).\\
  \end{array}
  \right.
\]

Then,  at the step $n = 1$, the following conditions are satisfied.
Note that there is an exchange of suffices of $\Psi_i$.

\[
\left\{
  \begin{array}{ll}

  f_1(\Psi_0) \subset \Omega_{-(a-b)}, & \\
  f_1(\Psi_1) \subset \Omega_{a+b}, &\\
                               &    \\
  f_1(X_0)    \subset \Psi_0,   &\\
  f_1(X_1)    \subset \Psi_1.   &\\

  \end{array}
  \right.
\]

At the next step, the $f_1$ evolves to the following $f_2$;
\[
\left\{
  \begin{array}{lllrl}

  f_2|_{\Psi_0} &= 
(1-\e)(-a - Eb) + \e (-(a-b)) &=& -a +Eb &= f_0|_{\Psi_0},\\
  f_2|_{\Psi_1} &= 
(1-\e)(a - Eb) + \e (a+b)     &=& a + Eb & =f_0|_{\Psi_1},\\
                               &    &\\
  f_2|_{X_0}    &= 
(1-\e) E(a+Eb) + \e (-a-Eb)   &=& -E(a+Eb)& = f_0|_{X_0},\\
  f_2|_{X_1}    &= 
(1-\e) (-E(a-Eb)) + \e (a - Eb) &=& E(a-Eb) & = f_0|_{X_1}.\\

  \end{array}
  \right.
\]

This $f_2$ coincides with $f_0$.  Hence $f_n$ is a period-2 function.
These dynamics are shown in Figure \ref{meta1}, while
Figure \ref{metasc1} shows a schematic representation of the dynamics.
Each arrow $A \rightarrow B$ in the figure indicates that $f_n(A)\subset B$.
$f_{\infty}(\Omega_i)$ is always included in $\Fix (f_{\infty})$ and 
$f_n(\Psi_0)$ is included in $\Omega_{-1}$ or $\Omega_{-\e}$ in turns.


\begin{figure}[htbp]
 \begin{center}
    \includegraphics[width=12cm,height=9cm]{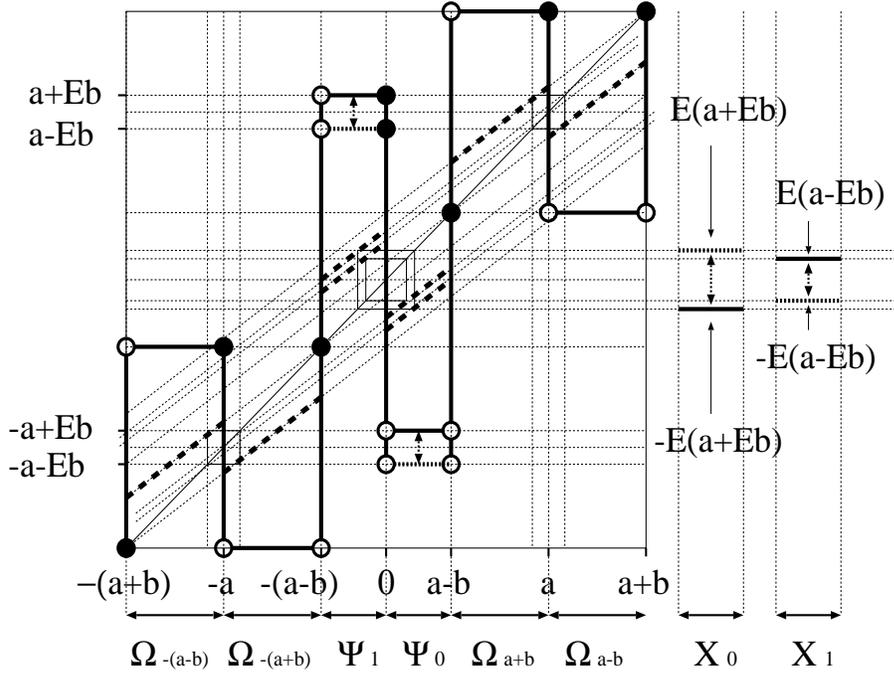}
 \end{center}
 \caption{\label{meta1} Temporal evolution of $f_n$ with a meta-map.
Here, $E := \frac{\e}{2-\e}$ and $b = \frac{1-\e}{1+\e}\cdot a$.
$f_n(x)$ at even step $n$ is plotted by solid line,
while that for odd $n$ is plotted by dotted line.
$g_n(x)$ is shown by bold dotted line.
$\e$ is set to $1/4$.
}
\end{figure}


\begin{figure}[htbp]
 \begin{center}
    \includegraphics[width=4cm,height=3cm]{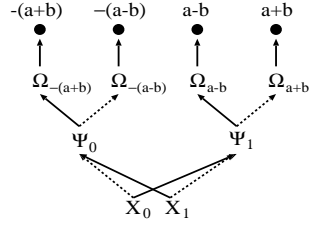}
 \end{center}
 \caption{\label{metasc1} The schematic representation of the dynamics.
The arrow $A\rightarrow B$ means that $f_n(A)\subset B$.
$f_{even}$ (solid line) and $f_{odd}$ (dotted line) are shown. 
The arrows over time steps (i.e., over the periods (=2))
are overlaid.}
\end{figure}

\end{example}

In the previous example,
the function which generates a map having a period-2 attractor
is embedded to give a new initial function $f_0$.
Note that the `surgery' of the initial function is valid
so that the generated map of the function has an arbitrary period.

The next example shows a meta-meta-map given by nesting this initial function.

\begin{example}[meta-meta-map]
\label{ex:5}
Define a new initial function $f_0$ by a recursive ``surgery'' of
the $f_0$ in Example \ref{ex:4} (meta-map).
The meta-meta-map is given by this recursive surgery.
In Figure \ref{metameta}, the hierarchical configuration of $f_0$ is plotted.
Two copies of the initial function in the Example \ref{ex:4} (meta-map) are embedded on the 
intervals $[-(a+b), -(a-b)]$ and $[a-b, a+b]$, for this new $f_0$.
Now, $f_0$ has 8 fixed points 
$\Fix (f_{\infty}) = \{\pm (a+b), \pm (a-b), a \pm c, -a \pm c\}$.
Here, $c := \frac{a-b}{a+b}\cdot b$.
According to the previous example, 
(i) each $g_n|_{\Phi}$ (meta-map) has period-2 attractors and  
(ii) $f_n|_X $ is arranged on the attractor of $g_n|_{\Psi}$. 
(iii) In this example, $f_n|_{X}$ is set 
to generate a time-dependent $g_n|_{X}$ which has period-2 attractors (meta-meta map).
(iv) Each $f_n|_{\Phi_i}$ ($f_n|_{\Phi_i} \subset X$ for all $n$) is on the attractor of $g_n|_{X}$
and gives a period-2 function.

Figure \ref{metametasc} shows a schematic representation of this case.
(i) $f_n|_{\Psi}$ is driven by $g|_{\Omega}$,
(ii) $f_n|_{X}$ driven by $g_n|_{\Psi}$ and
(iii) $f_n|_{\Phi}$ is driven by $g_n|_{X}$ hierarchically.
As is shown in this figure, one more step ($f_n|_{\Phi}$) is added
to the hierarchy in the Example \ref{ex:4} (Figure \ref{metasc1}), here.


\begin{figure}[htbp]
 \begin{center}
    \includegraphics[width=9cm,height=8cm]{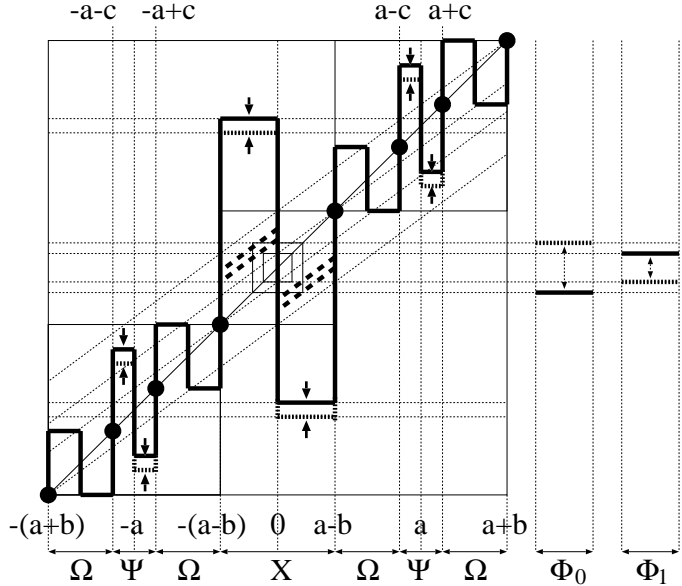}
 \end{center}
 \caption{\label{metameta} The evolution of $f_n$.  For the initial function,
two copies of the Example \ref{ex:4} (Figure \ref{meta1}) are embedded to $[\pm (a+b), \pm (a-b)]$.
Here, $\e$ is set to $1/4$ and $c := \frac{a-b}{a+b}\cdot b$.
}

\end{figure}


\begin{figure}[htbp]
 \begin{center}
    \includegraphics[width=8cm,height=5cm]{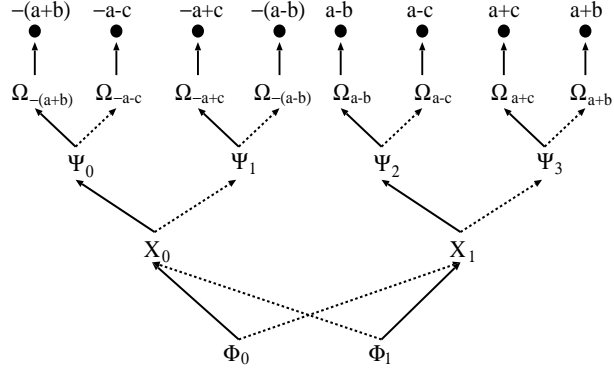}
 \end{center}
 \caption{\label{metametasc} The schematic representation of the dynamics. 
One more step ($f_n|_{\Phi}$) is added to the hierarchy in the Example \ref{ex:4} (Figure \ref{metasc1}).}
\end{figure}

\end{example}

As is described in the remark \ref{remarkhi}, 
this process can be continued ad infinitum.
A simple method to give an initial function with a higher hierarchical structure 
is to nest a given $f_0$ so that it satisfies  the condition for the
(extended) theorem \ref{theorem:3}.

\section{Further Example}

In the Examples \ref{ex:4} (meta-map) and \ref{ex:5} (meta-meta-map), the intervals are partially ordered at each step,
if the order is defined so that  $I_a < I_b$ iff $I_a \subset f_n(I_b)$ are satisfied 
(See Figure \ref{metasc1} and \ref{metametasc}).
In the Figures \ref{metasc1} and \ref{metametasc}, the arrows $A \rightarrow B$
for $f_n(A) \subset B$ change in time and the
arrows over time steps (i.e., over the periods (=2))
are overlaid.  Note that the intervals there are
partially ordered.
Generally, the intervals are not partially ordered for overlaid graph over $n$.
An example of the initial function for such case is given  below.
In this example, the hierarchy is ``entangled''.
There are some partial functions driven by each other generated map in turns.

\begin{example}[entangled hierarchy]
\label{ex:6}
Let $f_0(x)$ be as follows;
\[
  f_0(x) =
  \left\{
  \begin{array}{cll}
   \e-3,    & x \in \Omega_{\e-3},   &\Omega_{\e-3}:= [\e-3, -2]\cup (-1,0],\\
   3-\e,    & x \in \Omega_{3-\e},   &\Omega_{3-\e}:= (0, 1]\cup (2, 3-\e],\\
       & &\\
   1+\e, & x\in \Psi_0, &\Psi_0:= (-2, -1],\\
  -(1-\e), & x\in \Psi_1, &\Psi_1:= (1,2].\\
  \end{array}
  \right.
\]

In this example, the time evolution of this $f_0$ is demonstrated directly as follows.

Now, the initial function satisfies the following condition.
\[
\left\{
  \begin{array}{ll}

  f_0(\Psi_0) \subset \Psi_1, & \\
  f_0(\Psi_1) \subset \Omega_{\e-3}. &\\

  \end{array}
  \right.
\]

This $f_0$ evolves to the following $f_1$.
\[
\left\{
  \begin{array}{lll}

  f_1|_{\Psi_0} &= (1-\e) f_0|_{\Psi_0} + \e f_0|_{\Psi_1} \circ f_0|_{\Psi_0}& = 1-\e,\\
  f_1|_{\Psi_1} &= (1-\e) f_0|_{\Psi_1} + \e f_0|_{\Omega_{\e-3}} \circ f_0|_{\Psi_1}& = -(1+\e).\\

  \end{array}
  \right.
\]

At $n=1$ the relation
\[
\left\{
  \begin{array}{ll}

  f_1(\Psi_0) \subset \Omega_{3-\e}, & \\
  f_1(\Psi_1) \subset \Psi_0 &\\

  \end{array}
  \right.
\]
is satisfied. In this case, $\Psi$ is not only mapped to $\Omega$, but
also to $\Psi$ itself.  This $f_1$ evolves to the following $f_2$
\[
\left\{
  \begin{array}{lll}

  f_2|_{\Psi_0} &= (1-\e) f_1|_{\Psi_0} + \e f_1|_{\Omega_{3-\e}} \circ f_1|_{\Psi_0}& = f_0|_{\Psi_0},\\
  f_2|_{\Psi_1} &= (1-\e) f_1|_{\Psi_1} + \e f_1|_{\Psi_0}        \circ f_1|_{\Psi_1}& = f_0|_{\Psi_1}.\\

  \end{array}
  \right.
\]
This $f_2$ coincides with $f_0$.  Hence $f_n$ is a period-2 function.
These are shown in Figure \ref{meta2} while the
schematic representation is shown in Figure \ref{metasc2}.


\begin{figure}[htbp]
 \begin{center}
    \includegraphics[width=8cm,height=8cm]{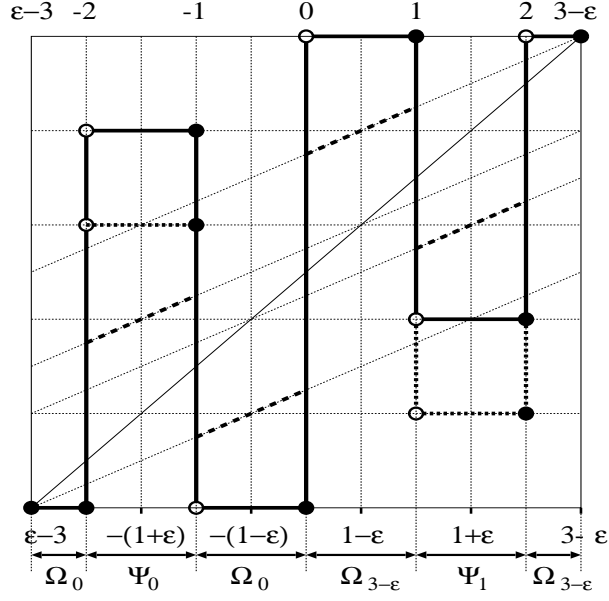}
 \end{center}
 \caption{\label{meta2} The evolution of the function with entangled hierarchy.
The solid line shows $f_{even}(x)$ and dotted line shows $f_{odd}(x)$.
The bold dotted line shows $g_n(x)$. 
$f_n|_{\Psi_i}$ evolves with period 2.
Here, $\e = 1/2$.}
\end{figure}


\begin{figure}[htbp]
 \begin{center}
    \includegraphics[width=4cm,height=3cm]{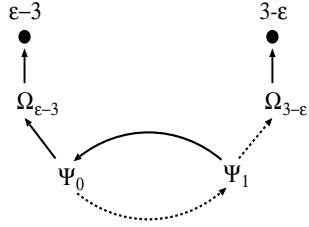}
 \end{center}
 \caption{\label{metasc2} The schematic representation of the ``entangled hierarchy''.
The arrows over time steps (i.e., over the periods (=2))
are overlaid.
There is a loop in this graph which does not exist in previous examples.}
\end{figure}

\end{example}

The loop in Figure \ref{metasc2} shows that the dynamics of $f_n|_{\Psi_0}$ and $f_n|_{\Psi_1}$ are determined by
$g_n|_{\Psi_1}$ and $g_n|_{\Psi_0}$ in turns.
Note that the dynamics of $f_n|_{\Psi_0}$ and $f_n|_{\Psi_1}$ are `not' determined each other at the same step $n$ in the Example \ref{ex:6}. 
The snapshot of the graph at $n$ is partially ordered,
while the overlaid graph for $n$ has the loop.

In Section \ref{sec:3}, all intervals are partially ordered.
There, the dynamics of $f_n|_{A}$ have no influence to the dynamics of $f_n|_{B}$, if $f_n(A)\subset B$.
Now the ``entanglement'' exists and the dynamics of $f_n|_A$ has the influence to $f_{m}|_B$ ($n\neq m$), 
even if the condition $f_n(A)\subset B$ are satisfied at some $n$.

\section{Discussion}
To close the paper, we briefly discuss the original motivation in
the study of (1.1) [1] and possible 
relevance of our result to a biological system.
In a biological system, we are often amazed at its ability to change its
own rule, while in a mechanical system there usually exists a rigid rule
which governs the change of the state forever.
Moreover, the rule in a biological system is 
formed `spontaneously', depending on the history of the state,
without being prescribed externally.  
There a rule to drive the change of the state and the state driven by
the rule are not separated initially, but through 
dynamics, some part of the
system starts to drive other parts, and works as a rule.

When we adopt usual dynamical systems on phase spaces, however,
the question how a rule is formed is not
answered, since in dynamical systems, the rule for
dynamics, and the variables that are driven by the rule are
clearly separated.  When a rule is not separated from the state, however,
the rule (that is undifferentiated from the state variables) may operate to itself. 
In our function dynamics, we
try to answer the problem of this self-operation of a rule
by explicitly taking into account the
term $f\circ f$, since with this term, the
function $f$ to change a state value $x$ can also be a state value to be
changed by it.

This $f \circ f$ term leads to a self-reference, since
the evolution of the function $f_n(x)$ obeys  the generated map $g_n(x)$,
which itself refers to the function $f_{n-1}(x)$.
Indeed the importance of self-reference is generally
discussed in a biological problem.
In our cognition, for example,
external inputs are processed and are mapped to an output.
The output from this process influences 
our cognitive process itself.  If we regard this
cognitive process as a function from inputs to outputs,
this function changes in time following some self-reference,
through development of our cognition.
Our study of the function dynamics (\ref{eqn:1.1}) was originally introduced
as a toy model to study the dynamics with such self-reference \cite{KKI},
and was motivated
by the search for a novel class of phenomena in a system with
self-referential structure.

In the structure of Section 3, we have demonstrated
that evolution of some partial functions is driven by the
generated map of some other intervals 
hierarchically.  
The generated map of some intervals works as a `rule' to drive 
other intervals, although they are not initially prescribed as a
rule part in our model equation. These intervals to drive other parts are 
given by flat parts of $f_n(x)$.  
In fact, with temporal evolution of our function dynamics, 
the whole interval is partitioned into flat parts.

In a biological system,  rules are 
often formed  first by partition of continuous inputs into
discrete symbols, and these 
symbols provide a basis for a syntactic structure to drive other
parts.  This partition process is  called articulation
in our cognition and language
(for example, continuous spectrum of light is `articulated' into
a discrete set of colors).
As mentioned, this articulation process and the generation of
rules over the articulated symbols
are a general feature of our function dynamics.

In the function dynamics, the rule, i.e., the
generated map, can change in time, when
the driving by a generated map has a hierarchical structure as in Section 3.
In this sense, the hierarchy of a rule, a rule to change the rule, the further rule
to change it, ... is formed in the function dynamics.
Such hierarchy in the change of rules also reminds us of 
hierarchical structures ubiquitous in a biological system, and also
in our cognitive process.
Furthermore, in Section 4, we have found an example in which the hierarchy 
structure itself can change in time, where the separation of
rule and state formed through dynamics is partially destroyed.
The rules and hierarchy in a biological system  
have stability on one hand, and plasticity on the other hand.
In future, it will be important to analyze the stability of the structure
we found in the paper.\\



 The late Professor Masaya Yamaguti stressed the importance
of self-reference from early days.  He often mentioned his
interest in fractals in connection with the self-reference.
Hata and he
studied function equation leading to fractal \cite{YM}.
Furthermore he often discussed that mathematics dealing with
self-reference is necessary to  psychology, natural language,
and so forth.  It is to be regretted that we could not present
our paper while he lived.

\end{document}